\documentclass[preprint,showpacs,preprintnumbers,amsmath,amssymb,superscriptaddress]{revtex4}
\usepackage{graphicx}
\usepackage{dcolumn}
\usepackage{bm}
\usepackage{longtable}
\newcommand{\beqn}{\begin{eqnarray}}
\newcommand{\eeqn}{\end{eqnarray}}

\begin{document}

\title{Examining the chiral geometry in $^{104}$Rh and
$^{106}$Rh}\thanks{This work is partly supported by the National
Natural Science Foundation of China under Grant No. 10605001,
10435010, 10221003, and 10505002, the postdoctoral Science
Foundation of China No. 20060390371.}
\author{WANG Shou-Yu}
\affiliation{School of Physics, Peking University, Beijing 100871,
China}
\author{ZHANG Shuang-Quan}\thanks{e-mail: sqzhang@pku.edu.cn}
\affiliation{School of Physics, Peking University, Beijing 100871,
China} \affiliation{Institute of Theoretical Physics, Chinese
Academy of Science, Beijing 100080, China}
\author{QI Bin}
\affiliation{School of Physics, Peking University, Beijing 100871,
China}
\author{MENG Jie}\thanks{e-mail: mengj@pku.edu.cn}
\affiliation{School of Physics, Peking University, Beijing 100871,
China} \affiliation{Institute of Theoretical Physics, Chinese
Academy of Science, Beijing 100080, China}
\affiliation{Center of Theoretical Nuclear Physics, National Laboratory of \\
       Heavy Ion Accelerator, Lanzhou 730000,
China}

\begin{abstract}
The criteria for chiral doublet bands based on one particle and one
hole coupled to a triaxial rotor have been summarized. Two
representative cases in $A\sim100$ odd-odd nuclei, nearly degenerate
$\Delta I =1$ doublet bands in $^{104}$Rh and $^{106}$Rh are checked
against these chiral criteria. It is shown that $^{106}$Rh possesses
better chiral geometry than $^{104}$Rh, although the energy near
degeneracy is achieved in $^{104}$Rh in comparison with the constant
energy separation of doublet bands in $^{106}$Rh.

\end{abstract}

\maketitle

Chirality is a common property in chemistry, biology and particle
physics. In nuclear physics, Frauendorf and Meng~\cite{FM97} pointed
out that the rotation of triaxial nuclei may attain the chiral
character and give rise to pairs of identical $\Delta I =1$ bands
with the same parity --- chiral doublet bands. Extensive
experimental studies have been made to search for such bands. So
far, candidate chiral doublet bands have been proposed in quite a
number of odd-odd nuclei in the $A\sim130$ and $A\sim100$ mass
regions~\cite{Starosta01,Koike01,AA01,Hartley01,LiXF02,Koike03,GR03,
Wang06,CD04,JO04,Joshi05,Yong05}. However, in none of the reported cases,
the genuine degeneracy between the two chiral partner bands were
observed extended over a certain spin region. In most cases, the
energy separation between the two partner bands ranges from
$\sim200$ keV to $\sim400$ keV. The smallest level degeneracy is
observed in $^{134}$Pr and $^{104}$Rh, where the states at spins
$15^{+}$ and $17^{-}$ are separated in energies 36 keV and 2 keV,
respectively. Thus, $^{134}$Pr and $^{104}$Rh have been considered
as the best examples of chiral rotation in the $A\sim130$ and
$A\sim100$ mass regions. However, the recent lifetime experiment for
$^{134}$Pr gives different E2 transition rates for the partner
bands~\cite{TO06}, which is contradictory to the chiral
interpretation.

In the $^{134}$Pr nucleus, as it was already known even before the
lifetime measurement that the $B(M1)/B(E2)$ ratios were different
for the doublet bands, either the $B(M1)$ or $B(E2)$ values must be
different. In the $A\sim100$ mass region, although the lifetime
measurement is not available yet, the inband $B(M1)/B(E2)$ ratios
have been known for the doublet bands. Based on the known
$B(M1)/B(E2)$ ratios, we may deduce whether the $B(M1)$ and $B(E2)$
values are same or not for the doublet bands. Thus, it is feasible
to examine the existence of the chiral nuclei even if the
experimental lifetime data are not available yet. In this paper, the
criteria for chiral rotation will be summarized and discussed, and
applied to the candidate chiral bands in odd-odd nuclei $^{104,
106}$Rh. Compared with the other candidate chiral bands
$^{100}$Tc~\cite{Joshi05}, $^{106}$Ag~\cite{He06} reported in the
$A\sim100$ mass region, there are relatively rich data in nuclei
$^{104,106}$Rh for the present investigation.


Recently, based on one particle and one hole coupled to a triaxial
rotor, a set of experimental signatures have been suggested as
fingerprints of nuclear chirality, e.g., see
~\cite{FM97,Peng03a,Peng03b,Koike03,CD04,Koike04,Petrache06}. All
these signatures are necessary to identify the ideal chiral bands.
The first signature is that nearly degenerate levels of the same
spin and parity in two $\Delta I =1$ bands built on the same
single-particle configuration. It is the most essential consequence
of the spontaneous formation of chirality~\cite{FM97}. Secondly, the
energy staggering parameter $S(I)=[E(I)-E(I-1)]/2I$, is indicative
of the degree of mutual orthogonality of the three vectors involved.
For chiral nuclei it has been shown that the $S(I)$ values should
possess a smooth dependence with spin since the particle and hole
orbital angular momenta are both perpendicular to the core
rotation~\cite{CD04}. Thirdly, due to the restoration of the chiral
symmetry in the laboratory frame there are phase consequences for
the chiral wavefunctions resulting in $M1$ and $E2$ selection rules
which can manifest as $B(M1)/B(E2)$ ratios staggering as a function
of spin~\cite{Koike03}. The odd spin values are  higher than the
even spin values for chiral bands built on the configurations $\pi
h_{11/2}\otimes\nu h_{11/2}$ in nuclei with $A\sim130$ while it is
reversed for the $\pi g_{9/2}\otimes\nu h_{11/2}$ configuration in
nuclei with $A\sim100$. Another important signature on the
electromagnetic properties of chiral geometry is that there should
be very similar $B(M1)$ and $B(E2)$ transition strengthes between
the chiral partner states~\cite{TO06, Petrache06}. This leads to the
expectation that the $B(M1)/B(E2)$ ratios are very similar for the
partner bands. The fifth signature proposed by Petrache et
al~\cite{Petrache06} is the spin alignments between the doublet
bands should be very similar. The similar spin alignments of doublet
bands is the necessary condition for claiming the same single
particle configuration. Finally, in an ideal chiral symmetry
scenario, it is expected that interband $B(E2,I\rightarrow I-2)$
transitions disappear in the chiral region~\cite{Koike04}.

In short, the known criteria for chiral doublet bands to date can
be summarized as:
 (i) the nearly degeneracy of doublet bands;
 (ii) the spin independence of $S(I)$;
 (iii) the staggering of $B(M1)/B(E2)$ ratios,
 (iv) the identical $B(M1)$ and $B(E2)$ values,
 (v) identical spin alignments,
 (vi) the vanishing of the interband $E2$ transitions at high spin region.
 In the following, the candidate chiral bands in $^{104}$Rh and
$^{106}$Rh will be examined on the basis of the above-mentioned
chiral criteria.

In order to discuss and compare observed partner bands in
$^{104,106}$Rh~\cite{CD04,JO04} together with those in
$^{134}$Pr~\cite{TO06,Petrache96}, the excitation energy (upper
panel) and the energy staggering parameter $S(I)=[E(I)-E(I-1)]/2I$
(lower panel), the intraband $B(M1)/B(E2)$, and the spin alignments
for the partner bands are plotted in Figs. 1, 2 and 3, respectively.
In fact, these figures or data are already available in the
literatures ( see the corresponding captions for the details ), and
they are summarized here again for the convenience of discussions.

Recent experimental efforts have led to the identification of the
doublet bands in $^{104}$Rh~\cite{CD04} and $^{106}$Rh~\cite{JO04}
based on the $\pi g_{9/2}\otimes\nu h_{11/2}$ configuration in the
100 mass region. As shown in Figs.1 (a, d) and 2 (a), the
experimental results in $^{104}$Rh show excellent agreement with
first three chiral criteria which have been previously
proposed~\cite{CD04}, namely, nearly degenerate doublet bands,
constant $S(I)$ as a function of spin and staggering of
$B(M1)/B(E2)$ ratios. Furthermore, the small energy separation of
the two bands in $^{104}$Rh($\sim2$ keV at spin $17^{-}$) provides
the best example of degeneracy. Thus, $^{104}$Rh has been considered
as the best candidate for chiral nuclei observed until now. However,
we note that the intraband $B(M1)/B(E2)$ ratios are appreciably
different between the two bands. As shown in fig.2(a, d), the
$B(M1)/B(E2)$ values for main band are a factor $\sim 2$ larger than
those of side band. This clearly fails to meet the fourth criterion.
In addition, the alignments of the side band are larger by
$\sim$2$\hbar$ than those of main band in a wide frequency interval,
as shown in fig.3(a). In the ideal chiral doublet bands, it is
expected that the interband $B(E2,I\rightarrow I-2)$ should be
vanishingly small at high spin. Whereas, the comparable strength of
interband and intraband $B(E2)$ transitions is observed for the
doublet bands in the whole spin range. These can be seen from Fig. 2
of Ref. \cite{CD04}. These experimental observations do not meet the
chiral criteria (4),(5) and (6). Thus, it is not in favour for
$^{104}$Rh to be considered as a good example of the chirality. In
order to further check for the chiral geometry in $^{104}$Rh, the
lifetime measurements of the doublet bands are highly suggested to
be performed.

The $\pi g_{9/2}\otimes\nu h_{11/2}$ doublet band properties of
$^{106}$Rh are shown in figures 1(b, e), 2(b) and 3(b). The energy
differences, as shown in figure 1(b), are rather constant around
300 keV up to high spin states. The overall $S(I)$ values are
close to constant without any clear zig-zag pattern for both
bands(see figure 1(e)). Compared to $^{104}$Rh, the intraband
$B(M1)/B(E2)$ ratios for the two bands are close to each other
over a large spin range, except for I=12 $\hbar$, and show
odd-even staggering as function of spin. The alignments of the
main and side band overlap in a wide frequency interval before
bandcrossing. These can be seen in Fig.2(b, e) and 3(b),
respectively. In additional, as shown in Fig. 1 of Ref.
\cite{JO04}, none E2 linking transitions between main band and
side band are observed in the chiral region, which is consistent
with the ideal chiral geometry. Based on the above arguments, all
available experimental results on the doublet bands of $^{106}$Rh
to date well agreed with the expected chiral criteria. As shown in
figures 1, 2, 3 and Table I, $^{106}$Rh meets more the ideal
chiral criteria than $^{104}$Rh. In terms of the consistency with
the expected chiral characteristics, the constant energy
separation in doublet bands of $^{106}$Rh possesses better chiral
geometry than the achieving energy degeneracy in doublet bands of
$^{104}$Rh. It is consistent with the potential energy
surface~\cite{JO04} and microscopic relativistic mean
field~\cite{JM06} calculations. These calculations indicate that
$^{106}$Rh could provide a better example of chiral geometry than
that found in $^{104}$Rh, since the triaxial minimum for the $\pi
g_{9/2}\otimes\nu h_{11/2}$ configuration in $^{106}$Rh is
predicted to be much closer to $30^{\circ}$ than in $^{104}$Rh.
Moreover, the present analysis further indicates that the
achieving energy degeneracy is not necessarily indicative of a
better chiral bands than the constant energy separation of doublet
bands. In an ideal chiral symmetry scenario, the doublet bands are
expected to be degenerate over a certain spin range. However, the
nucleus is seldom an ideal system, and the actual situation is
expected to be considerably more complicated and dynamic. The
constant energy separation in the certain spin region can be
understood as that there is the constant barrier of quantum
tunnelling between the left and right handed systems.

Most of the experimental results, on which the doublet bands in
$^{104}$Rh are not consistent with the chiral geometry, are also
present in the doublet bands of $^{134}$Pr. These common features
can be summarized as following: (1), the $B(M1)/B(E2)$ ratios in
the side band are apparently larger than that in the main band at
the whole spin region. (2), The alignments of the side band are
larger by $\sim$2$\hbar$ than those of the main band in a wide
frequency interval. (3), The nonzero values of interband
$B(E2,I\rightarrow I-2)$ is observed in the whole spin region. On
the other hand, a notable phenomenon is that the $E(I)$ plots of
the doublet bands in $^{104}$Rh and $^{134}$Pr exhibit very
similar shapes, seen from Fig. 1 (a, c). The similar relative
energy displacement, the alignments and electromagnetic properties
observed in $^{104}$Rh and $^{134}$Pr implies similar underlying
mechanisms for the band doublings in the two nuclei.

In Ref.~\cite{Petrache06}, the measured B(E2) values within the
doublet bands in $^{134}$Pr are not identical while the
corresponding B(M1) values have a similar behaviour within the
experimental uncertainties. So the difference of intraband
$B(M1)/B(E2)$ ratios
 may be only attributed to the difference of $B(E2)$
values. Similar to $^{134}$Pr, if only the $B(E2)$ values
difference are taken into account for offering the intraband
$B(M1)/B(E2)$ ratios difference within the two bands in
$^{104}$Rh, we can obtain that the $B(E2)$ values for main band
are a factor 2 larger than those of side band, since the identical
$B(M1)$ strengths is assumed. In the rotational
model~\cite{Bohr75}, the $B(E2)$ values can be related to the
transition quadrupole moments:
 \begin{eqnarray}
 B(E2;I_{i}K\rightarrow I_{f}K)
 = \frac{5}{16\pi}e^{2}Q_{0}^{2}\langle I_{i}K20|I_{f}K\rangle^2
 \end{eqnarray}
Therefore, $Q_{0,main}/Q_{0,side}\simeq1.4$ can be obtained for
the doublet bands in $^{104}$Rh. It indicates that the shapes of
nearly degeneracy doublet bands are very different. Thus the shape
coexistence may be suggested for two bands having the same orbital
configuration. In order to clarify the mechanism for the partner
bands, further theoretical and experimental studies are necessary,
such as the lifetime measurement.

In summary, The criteria for chiral doublet bands based on one
particle and one hole coupled to a triaxial rotor have been
summarized. The experimental features of the doublet bands in two
representative cases in $A\sim100$ mass region, $^{104}$Rh and
$^{106}$Rh are checked against the configuration-independent chiral
criteria. The present analysis indicates that $^{106}$Rh possesses
better chiral geometry than $^{104}$Rh, although the energy near
degeneracy is achieved in $^{104}$Rh in comparison with the constant
energy separation of doublet bands in $^{106}$Rh.

\vspace{1.0cm}
\renewcommand\refname{References}
\renewcommand{\baselinestretch}{1.5}

\newpage
\begin{table*}[htbp]\centering
    \setlength{\tabcolsep}{1.2em}
    \caption{Check for the chiral geometry in $^{104}$Rh , $^{106}$Rh and $^{134}$Pr with the ideal chiral criteria.}\label{Tab1-1}
    \begin{tabular}{c|rrr}
        \hline\hline
            Ideal chiral criteria                & $^{104}$Rh & $^{106}$Rh & $^{134}$Pr \\ \hline
        Nearly degeneracy doublet bands          & $\surd$  & $\surd$  & $\surd$  \\
        $S(I)$ independent of spin               & $\surd$  & $\surd$  & $\surd$ \\
        $B(M1)/B(E2)$ staggering                 & $\surd$  & $\surd$  & $\times$ \\
        Identical $B(M1)/B(E2)$                  & $\times$ & $\surd$  & $\times$ \\
        Identical spin alignments                & $\times$ & $\surd$  & $\times$ \\
        $B(E2)_{out}$ vanishing at high spin     & $\times$ & $\surd$  & $\times$ \\
        \hline\hline
    \end{tabular}
\end{table*}

\begin{figure*}
\includegraphics[bb=20 135 440 450,scale=0.8]{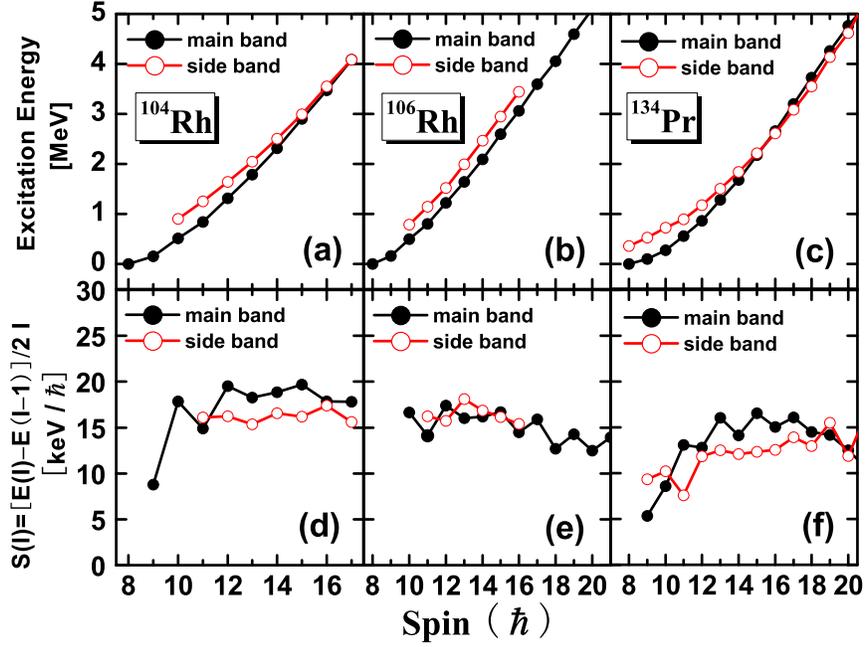}
\caption{\label{fig:epsart}Excitation energies relative to a
rigid-rotor reference (upper panel) and $S(I)$ values vs spin for
the partner bands in $^{104}$Rh, $^{106}$Rh and $^{134}$Pr. The data
are respectively from Ref. ~\cite{CD04}, \cite{JO04} and
\cite{TO06}. The J parameters are evaluated from the relation
$J=0.007\times(\frac{158}{A})^{5/3}$ MeV.}
\end{figure*}

\begin{figure*}
\includegraphics[bb=35 28 593 432,scale=0.6]{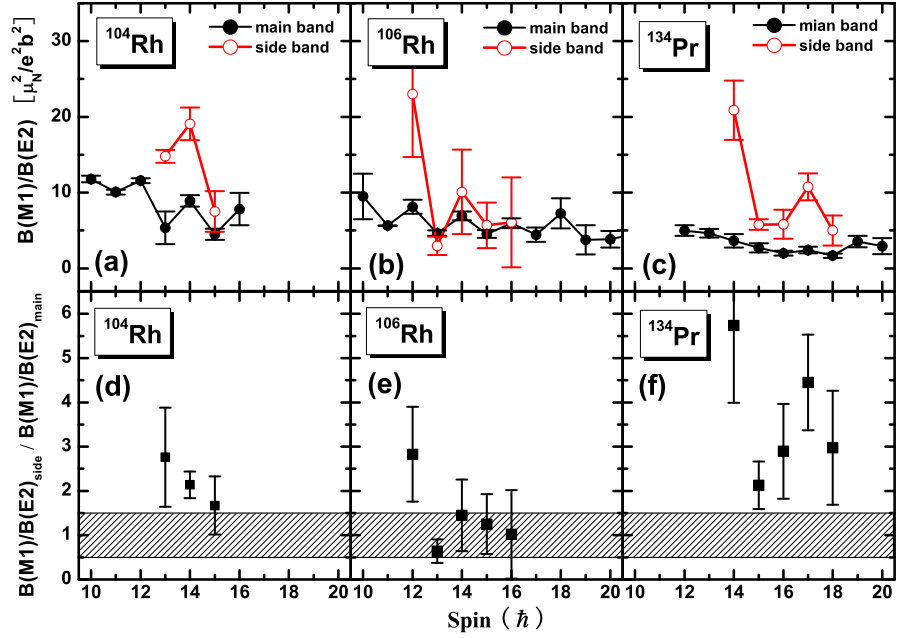}
\caption{\label{fig:epsart}The intraband $B(M1)/B(E2)$ (upper panel)
and ratio of $B(M1)/B(E2)$ values for the partner bands in
$^{104,106}$Rh and $^{134}$Pr. The data are respectively from Ref.
~\cite{CD04}, \cite{JO04} and \cite{Petrache96}.}
\end{figure*}

\begin{figure*}
\includegraphics[bb=15 15 320 120,scale=1.5]{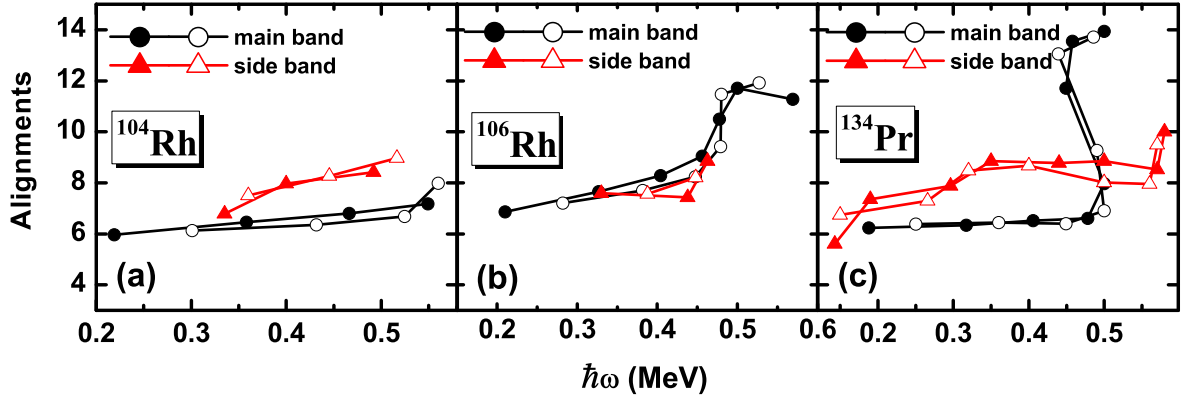}
\caption{\label{fig:epsart} Rotational alignments for the doublet
bands in $^{104}$Rh, $^{106}$Rh and $^{134}$Pr. Circles denote the
main bands, triangles denote the side bands. The data are
respectively from Ref. ~\cite{CD04}, \cite{JO04} and
\cite{TO06}.}
\end{figure*}

\end{document}